\documentclass[pra,aps,twocolumn,showpacs]{revtex4}

\usepackage{amsmath,amsfonts,amssymb,caption,hyperref,color,epsfig,graphics,graphicx,latexsym,mathrsfs,revsymb,theorem,url,verbatim,epstopdf}

\hypersetup{colorlinks,linkcolor={blue},citecolor={blue},urlcolor={red}}

\usepackage{hyperref}

\makeatletter

\newcommand{\Rmnum}[1]{\expandafter\@slowromancap\romannumeral #1@}
\makeatother
\newtheorem{definition}{Definition}
\newtheorem{proposition}[definition]{Proposition}
\newtheorem{Lemma}[definition]{Lemma}

\newtheorem{Theorem}[definition]{Theorem}

\newtheorem{conjecture}[definition]{Conjecture}

\newtheorem{remark}[definition]{Remark}
\newtheorem{example}[definition]{Example}
\newtheorem{question}[definition]{Question}

\def\squareforqed{\hbox{\rlap{$\sqcap$}$\sqcup$}}
\def\qed{\ifmmode\squareforqed\else{\unskip\nobreak\hfil
		\penalty50\hskip1em\null\nobreak\hfil\squareforqed
		\parfillskip=0pt\finalhyphendemerits=0\endgraf}\fi}
\def\endenv{\ifmmode\;\else{\unskip\nobreak\hfil
		\penalty50\hskip1em\null\nobreak\hfil\;
		\parfillskip=0pt\finalhyphendemerits=0\endgraf}\fi}
\newenvironment{proof}{\noindent \textbf{{Proof.~} }}{\qed}
\def\Dbar{\leavevmode\lower.6ex\hbox to 0pt
	{\hskip-.23ex\accent"16\hss}D}
\makeatletter
\def\url@leostyle{%
	\@ifundefined{selectfont}{\def\UrlFont{\sf}}{\def\UrlFont{\small\ttfamily}}}
\makeatother
\def\bcj{\begin{conjecture}}
	\def\ecj{\end{conjecture}}
\def\bcr{\begin{corollary}}
	\def\ecr{\end{corollary}}
\def\bd{\begin{definition}}
	\def\ed{\end{definition}}
\def\bea{\begin{eqnarray}}
\def\eea{\end{eqnarray}}
\def\bem{\begin{enumerate}}
	\def\eem{\end{enumerate}}
\def\bex{\begin{example}}
	\def\eex{\end{example}}
\def\bim{\begin{itemize}}
	\def\eim{\end{itemize}}
\def\bl{\begin{lemma}}
	\def\el{\end{lemma}}
\def\bma{\begin{bmatrix}}
	\def\ema{\end{bmatrix}}
\def\bpf{\begin{proof}}
	\def\epf{\end{proof}}
\def\bpp{\begin{proposition}}
	\def\epp{\end{proposition}}
\def\bqu{\begin{question}}
	\def\equ{\end{question}}
\def\br{\begin{remark}}
	\def\er{\end{remark}}
\def\bt{\begin{theorem}}
	\def\et{\end{theorem}}

\def\btb{\begin{tabular}}
	\def\etb{\end{tabular}}

\newcommand{\nc}{\newcommand}

\def\a{\alpha}
\def\b{\beta}

\def\r{\rho}
\def\s{\sigma}

\def\o{\omega}

\nc{\bbA}{\mathbb{A}} \nc{\bbB}{\mathbb{B}} \nc{\bbC}{\mathbb{C}}
\nc{\bbD}{\mathbb{D}} \nc{\bbE}{\mathbb{E}} \nc{\bbF}{\mathbb{F}}
\nc{\bbG}{\mathbb{G}} \nc{\bbH}{\mathbb{H}} \nc{\bbI}{\mathbb{I}}
\nc{\bbJ}{\mathbb{J}} \nc{\bbK}{\mathbb{K}} \nc{\bbL}{\mathbb{L}}
\nc{\bbM}{\mathbb{M}} \nc{\bbN}{\mathbb{N}} \nc{\bbO}{\mathbb{O}}
\nc{\bbP}{\mathbb{P}} \nc{\bbQ}{\mathbb{Q}} \nc{\bbR}{\mathbb{R}}
\nc{\bbS}{\mathbb{S}} \nc{\bbT}{\mathbb{T}} \nc{\bbU}{\mathbb{U}}
\nc{\bbV}{\mathbb{V}} \nc{\bbW}{\mathbb{W}} \nc{\bbX}{\mathbb{X}}
\nc{\bbZ}{\mathbb{Z}}

\nc{\bA}{{\bf A}} \nc{\bB}{{\bf B}} \nc{\bC}{{\bf C}}
\nc{\bD}{{\bf D}} \nc{\bE}{{\bf E}} \nc{\bF}{{\bf F}}
\nc{\bG}{{\bf G}} \nc{\bH}{{\bf H}} \nc{\bI}{{\bf I}}
\nc{\bJ}{{\bf J}} \nc{\bK}{{\bf K}} \nc{\bL}{{\bf L}}
\nc{\bM}{{\bf M}} \nc{\bN}{{\bf N}} \nc{\bO}{{\bf O}}
\nc{\bP}{{\bf P}} \nc{\bQ}{{\bf Q}} \nc{\bR}{{\bf R}}
\nc{\bS}{{\bf S}} \nc{\bT}{{\bf T}} \nc{\bU}{{\bf U}}
\nc{\bV}{{\bf V}} \nc{\bW}{{\bf W}} \nc{\bX}{{\bf X}}
\nc{\bZ}{{\bf Z}}

\nc{\cA}{{\cal A}} \nc{\cB}{{\cal B}} \nc{\cC}{{\cal C}}
\nc{\cD}{{\cal D}} \nc{\cE}{{\cal E}} \nc{\cF}{{\cal F}}
\nc{\cG}{{\cal G}} \nc{\cH}{{\cal H}} \nc{\cI}{{\cal I}}
\nc{\cJ}{{\cal J}} \nc{\cK}{{\cal K}} \nc{\cL}{{\cal L}}
\nc{\cM}{{\cal M}} \nc{\cN}{{\cal N}} \nc{\cO}{{\cal O}}
\nc{\cP}{{\cal P}} \nc{\cQ}{{\cal Q}} \nc{\cR}{{\cal R}}
\nc{\cS}{{\cal S}} \nc{\cT}{{\cal T}} \nc{\cU}{{\cal U}}
\nc{\cV}{{\cal V}} \nc{\cW}{{\cal W}} \nc{\cX}{{\cal X}}
\nc{\cZ}{{\cal Z}}

\nc{\hA}{{\hat{A}}} \nc{\hB}{{\hat{B}}} \nc{\hC}{{\hat{C}}}
\nc{\hD}{{\hat{D}}} \nc{\hE}{{\hat{E}}} \nc{\hF}{{\hat{F}}}
\nc{\hG}{{\hat{G}}} \nc{\hH}{{\hat{H}}} \nc{\hI}{{\hat{I}}}
\nc{\hJ}{{\hat{J}}} \nc{\hK}{{\hat{K}}} \nc{\hL}{{\hat{L}}}
\nc{\hM}{{\hat{M}}} \nc{\hN}{{\hat{N}}} \nc{\hO}{{\hat{O}}}
\nc{\hP}{{\hat{P}}} \nc{\hR}{{\hat{R}}} \nc{\hS}{{\hat{S}}}
\nc{\hT}{{\hat{T}}} \nc{\hU}{{\hat{U}}} \nc{\hV}{{\hat{V}}}
\nc{\hW}{{\hat{W}}} \nc{\hX}{{\hat{X}}} \nc{\hZ}{{\hat{Z}}}

\nc{\hn}{{\hat{n}}}

\def\tr{\mathrm{tr}}

\def\max{\mathop{\rm max}}
\def\min{\mathop{\rm min}}

\def\tr{\mathop{\rm Tr}}

\newcommand{\bra}[1]{\langle#1|}
\newcommand{\ket}[1]{|#1\rangle}

\newcommand{\norm}[1]{\lVert#1\rVert}

\def\Dbar{\leavevmode\lower.6ex\hbox to 0pt
	{\hskip-.23ex\accent"16\hss}D}

\begin{document}
	\title{Lower Bounds of Entanglement Quantifiers Based On Entanglement Witnesses}
	
	\author{Xian Shi}\email[]
	{shixian01@gmail.com}
	\affiliation{College of Information Science and Technology,
		Beijing University of Chemical Technology, Beijing 100029, China}

	%
	
	
	
	\date{\today}
	
	
	\begin{abstract}
	To quantify the entanglement of bipartite systems in terms of some entanglement measure is a challenging problem in general, and it is much worse when the information about the system is less. In this manuscript, based on two classes of entanglement criteria, we present a method to obtain the lower bounds of the entanglement measures, concurrence, entanglement of formation, and geometrical entanglement measure. 
 	\end{abstract}
 \maketitle
	\section{Introduction}
	\indent	Entanglement is one of the essential features in quantum mechanics when compared with the classical physics \cite{horodecki2009quantum,plenio2014introduction}. It plays key roles in quantum information processing, such as quantum cryptography \cite{ekert1991quantum}, teleportation \cite{bennett1993teleporting}, and superdense coding \cite{bennett1992communication}. \par
	One of the most important problems in quantum information theory is how to distinguish separable and entangled states. The above problem is completely solved for $2\otimes2$ and $2\otimes3$ systems by the Peres-Horodecki criterion: a bipartite state $\r_{AB}$ is separable if and only if it is positive partial transpose (PPT), $i.$ $e.$, $(id\otimes T)(\rho_{AB})\ge 0$ \cite{peres1996separability}. However, the problem is NP-hard for arbitrary dimensional systems \cite{gurvits2003classical}. One of the most efficient tools to deal with the problem is entanglement witness \cite{chruscinski2014entanglement}. An entanglement witness is some observable that are positive on separable states while negative on some entanglement state. Recently, classes of entanglement witnesses have been built based on the properties of mutually unbiased bases \cite{chruscinski2018entanglement}, symmetric informationally complete measurements \cite{li2020entanglement}, and mutually unbiased measurements \cite{siudzinska2021entanglement}. Moreover, in the case of an ensemble with cold atoms \cite{hald1999spin} or trapped ions \cite{blatt2008entangled}, a collective of spin-squeezing inequalities was proposed to detect the entanglement between the particles \cite{sorensen2001many,vitagliano2011spin,vitagliano2014spin,Imai2023entanglement}. Besides, the entanglement between two spatially separated ensembles has been verified \cite{lange2018entanglement,shin2019bell,colciaghi2023einstein,eckart2023ultrafast}. Compared with the detection of entanglement of a bipartite system, the problem on how to quantify the entanglement is even harder. To deal with the problems, various entanglement measures have been proposed in the past years \cite{wootters1998entanglement,vedral1998entanglement,vidal1999robustness,terhal2000schmidt,vidal2000entanglement,wei2003geometric,christandl2004squashed,shi2021extension}. Among them, Concurrence \cite{wootters1998entanglement} and entanglement of formation \cite{bennett1996mixed} are two popular entanglement measures. However, these entanglement measures are difficult to compute for higher dimensional systems \cite{Huang_2014}, even if the whole information of bipartite states are obtained. Lots of results have been made on the lower bounds of concurrence and entanglement of formation \cite{chen2005concurrence,brandao2005quantifying,de2007lower,Chen2012,li2020improved,shi2023family}. Nevertheless, many schemes need to perform full tomography of the state to obtain the whole elements of the density matrices, that is, the entire process of getting the bounds of the entanglement is much more costly. Then, how to quantify the entanglement of an unknown state with fewer resources.

	In this manuscript, we present a method to evaluate the lower bounds of a bipartite system, which are based on the entanglement witnesses. The tool here we used is the Frobenius norm. Based on the properties of the norm, we present a lower bound of the distance between a bipartite mixed state and the set of separable states in terms of the Frobenius norm. Moreover, we also show the relations between the distance and the entanglement measures, concurrence, entanglement of formation and the geometrical entanglement measure. Hence, the lower bounds of the concurrence and entanglement of formation of the bipartite system are also obtained.
	
	This manuscript is organized as follows. In section \ref{se2}, we present the preliminary knowledge needed. In section \ref{se3}, we first present the lower bounds of the distance between a bipartite mixed state and the set of separable states based on two classes of entanglement criteria in terms of the Frobenius norm. Next, we show analytical expressions of the distance between a bipartite pure state and the set of separable states. At last, we obtain the lower bounds of the following entanglement measures: concurrence, entanglement of formation, and geometrical entanglement measure. In section \ref{se4}, we end with a conclusion.
	
\section{Preliminary Knowledge} \label{se2}
Here we first denote the following notations used in this manuscript. Let $\mathcal{L}(\mathcal{H})$, $\mathcal{L}_{+}(\mathcal{H})$, and $\mathcal{D}(\mathcal{H})$ be the sets consisting of linear operators, positive operators and states on a finite dimensional Hilbert space $\mathcal{H},$ respectively. Assume $A$ and $B$ are two operators in $\mathcal{L}(\mathcal{H}).$ We can endow $\mathcal{L}(\mathcal{H})$ with the following inner product between $A,B\in \mathcal{L}(\mathcal{H})$ 
\begin{align*}
\bra{A}B\rangle=\tr(A^{\dagger}B).
\end{align*}
The inner product can induce the Frobenius norm
\begin{align*}
\norm{A}_{F}=\sqrt{\bra{A}A\rangle}=\sqrt{A^{\dagger}A}.
\end{align*}

Assume $\mathcal{H}_{AB}=\mathcal{H}_A\otimes\mathcal{H}_B$ with $dim\mathcal{H}_A=d_A$ and $dim\mathcal{H}_B=d_B.$ If $\rho_{AB}$ is a state on $\mathcal{H}_{AB}$ that can be written as
 \begin{align*}
\r_{AB}=\sum_i p_i\rho_A^i\otimes\r_B^i,
\end{align*}
here $\{p_i\}_i$ is a probability distribution, $\rho_A^i$ and $\rho_B^i$ are states of subsystems $A$ and $B$, respectively, then it is separable. Otherwise, it is entangled. Below, we denote $Sep(A:B)$ as the set of separable states in $\mathcal{D}(\mathcal{H}_{AB}).$

Next we recall the entanglement measures considered here. Assume $\ket{\psi}_{AB}$ is a bipartite pure state. Due to the Schmidt decomposition, $\ket{\psi}_{AB}$ can always be written as $$\ket{\psi}_{AB}=\sum_i \sqrt{\lambda_i}\ket{i}_A\ket{i}_B,$$ where $\lambda_i\ge 0,$ $\sum_i\lambda_i=1,$ and $\{\ket{i}_{A(B)}\}$ is an orthonormal basis of the Hilbert space $A(B).$ The Entanglement of formation (EoF) of $\ket{\psi}_{AB}$ is given by
\begin{align}\label{Ep}
E(\ket{\psi}_{AB})=S(\rho_A)=-\sum \lambda_i\log_2\lambda_i,
\end{align} 
where $\lambda_i$ are the eigenvalues of $\rho_A=\tr_B\ket{\psi}_{AB}\bra{\psi}.$ For a mixed state $\rho_{AB},$ and EoF for a mixed state $\r_{AB}$ is defined by the convex roof extension method,
\begin{align}\label{Em}
E(\rho_{AB})=\min_{\{p_i,\ket{\phi_i}_{AB}\}}\sum_i p_i E(\ket{\phi_i}_{AB}).
\end{align} 
where the minimum takes over all the decompositions of $\rho_{AB}=\sum_i p_i\ket{\phi_i}_{AB}\bra{\phi_i}$, with $p_i\ge 0$ and $\sum p_i=1.$\\
\indent The other important entanglement measure is the concurrence ($C$). The concurrence of a pure state $\ket{\psi}_{AB}$ is defined as 
\begin{align}\label{Cp}
C(\ket{\psi}_{AB})=\sqrt{2(1-\tr\rho_A^2)}=\sqrt{2(1-\sum_i\lambda_i^2)}.
\end{align}
For a mixed state $\rho_{AB},$ it is defined as
\begin{align}\label{Cm}
C(\rho_{AB})=\min_{\{p_i,\ket{\phi_i}_{AB}\}}\sum_i p_i C(\ket{\phi_i}_{AB}),
\end{align}
where the minimum takes over all the decompositions of $\rho_{AB}=\sum_i p_i\ket{\phi_i}_{AB}\bra{\phi_i}$ with $p_i\ge 0$ and $\sum p_i=1.$ Another entanglement measure is the geometrical entanglement measure $E_g$. For a pure state $\ket{\psi}_{AB}=\sum_{i=0}^{d-1}\sqrt{\lambda_i}\ket{ii},$
\begin{align}
E_g(\ket{\psi}_{AB})=\min_{\s=\ket{\phi_1}_A\ket{\phi_2}_B}1-|\bra{\psi}\s\ket{\psi}|=1-\lambda_0.
\end{align} 
where the minimum takes over all the product state $\s.$

For a mixed state $\rho_{AB},$ it is defined as
\begin{align}
E_g(\rho_{AB})=\min_{\{p_i,\ket{\phi_i}_{AB}\}}\sum_i p_i C(\ket{\phi_i}_{AB}),
\end{align}
where the minimum takes over all the decompositions of $\rho_{AB}=\sum_i p_i\ket{\phi_i}_{AB}\bra{\phi_i}$ with $p_i\ge 0$ and $\sum p_i=1.$ 

At last, we recall a quantifier of bipartite states which denotes the distance to the set of separable states, $D_{sep}(\cdot)$, it is helpful to obtain the bounds of the above entanglement measures. Assume $\rho_{AB}$ is a mixed state, its distance to the set of separable states in term of $\norm{\cdot}_F$ is defined as
\begin{align*}
D_{sep}(\rho)=\min_{\s}\norm{\r-\s}_F,
\end{align*}
where the minimum takes over all the separable states $\s\in Sep(A:B)$. 
\section{Main Results}\label{se3}
In this manuscript, we mainly present the lower bounds of some entanglement measures of a given state when owing its knowledge obtained by classes of entanglement witnesses. Specifically,  First, we present the lower bounds of $D_{sep}(\cdot)$ based on classes of entanglement criteria. Next we bring the analytical formula of $D_{sep}(\cdot)$ for bipartite pure states, At last, by showing the relationship between $D_{sep}(\cdot)$ and some entanglement measures, we obtain the lower bounds of the entanglement measures.

\subsection{Lower bounds of $D_{sep}(\rho)$ derived from two entanglement criteria}
\indent In this subsection, we will present the bounds of $D_{sep}(\cdot)$ for a bipartite mixed state based on the entanglement witnesses obtained from mutually unbiased bases and spin-squeezing inequalities. 

First we consider a generic entanglement witness. Assume $W_0$ is an entanglement witness of $\mathcal{H}_d\otimes\mathcal{H}_d,$ let $a=\frac{\tr W_0}{d},$ $b=\sqrt{\tr(W_0^{\dagger}W_0)-\frac{(\tr W_0)^2}{d^2}},$ $W_1=\frac{W_0-aI\otimes I}{b}.$ 
Based on the property of $\norm{\cdot}_F$, we have
\begin{align}
D_{sep}(\rho_{AB})=&\min_{\s\in SEP(A:B}\max_{\norm{W}_F=1}\tr(W(\rho_{AB}-\s_{AB}))\nonumber\\
\ge &|\tr W_1(\rho-\o)]|\nonumber\\
=&|\tr(\frac{W_0}{b}(\rho-\o)-\frac{a}{b}\tr(\r-\o))|\nonumber\\
=&|\tr[\frac{W_0}{b}(\rho-\o)]|\nonumber\\
\ge&-\frac{1}{b}\tr W_0\r.\label{se}
\end{align}
In the first inequality, $\o$ is a separable state, $W_1$ is a Hermite operator and $\norm{W_1}_F=1.$ The last inequality is due to $\tr W_0\o\ge0.$

Next we recall the definitions of MUB. Let $\{\ket{\psi_k}\}$ and $\{\ket{\phi_l}\}_l$ be two orthonormal bases in $\mathcal{H}_d$. If for any $k,l$,
\begin{align*}
|\bra{\psi_k}\phi_l\rangle|=\frac{1}{d},
\end{align*}
then $\{\ket{\psi_k}\}_k$ and $\{\ket{\phi_l}\}_l$ are MUBs. Next we assume $\{\ket{\psi_1^{(\alpha)}},\ket{\psi_2^{(\a)}},\cdots,\ket{\psi_d^{\a}}\}$ with $\a=1,2\cdots,L$ are $L$ MUBs. In \cite{chruscinski2018entanglement}, the authors proposed the following entanglement witness 
\begin{align}
W_L=\frac{d-1+L}{d}I_d\otimes I_d-\sum_{\a=1}^{L}\sum_{k,l=1}^d\mathcal{O}_{kl}^{(\a)}\overline{P_l}^{(\a)}\otimes P_k^{(\a)},
\end{align}
here $P_k^{(\a)}=\ket{\psi_k^{(\a)}}\bra{\psi_k^{(\a)}}$, $\mathcal{O}^{(\alpha)}$ is a set of orthogonal rotation in $\mathbb{R}^d$ around the axis $\overrightarrow{\boldsymbol{n}}=\frac{1}{\sqrt{d}}(1,1,\cdots,1)$, that is, $\mathcal{O}^{(\a)}\overrightarrow{\boldsymbol{n}}=\overrightarrow{\boldsymbol{n}}.$ When $L=d+1$, $W$ can be written as
\begin{align}
W_{d+1}=2\mathbb{I}_d\otimes\mathbb{I}_d-\sum_{\a=1}^{d+1}\sum_{k,l=1}^d\mathcal{O}_{kl}^{(\a)}\overline{P_l}^{(\a)}\otimes P_k^{(\a)}.
\end{align}
Let $V_1=\sum_{\a=1}^{L}\sum_{k,l=1}^d\mathcal{O}_{kl}^{(\a)}\overline{P_l}^{(\a)}\otimes P_k^{(\a)},$ 
\begin{align*}
&\tr(V_1)\\
=&\tr[\sum_{\a=1}^{L}\sum_{k,l=1}^d\mathcal{O}_{kl}^{(\a)}\overline{P_l}^{(\a)}\otimes P_k^{(\a)}]\\
=&\tr[\sum_{\a=1}^{L}\sum_{k,l=1}^d\mathcal{O}_{kl}^{(\a)}\\
=&Ld\\
&\tr{V_1^{\dagger}V_1}\\
=&\tr[\sum_{\a=1}^{L}\sum_{k,l=1}^d\mathcal{O}_{kl}^{\a}\overline{P_l}^{(\a)}\otimes P_k^{(\a)}]^{\dagger}[\sum_{\b=1}^{L}\sum_{m,n=1}^d \mathcal{O}_{mn}\overline{P_n}^{\b}\otimes{P_m}^{(\b)}]\\
=&\tr[\sum_{\a,\b=1}^L\sum_{k,l,m,n=1}^d\mathcal{O}_{kl}^{(\a)}\mathcal{O}_{mn}^{(\b)}(\overline{P_l^{(\a)}})^{\dagger}\overline{P_n^{(\b)}}\otimes (P_k^{(\a)})^{\dagger}P_m^{\b}]\\
=&\sum_{\a\ne\b=1}^L\sum_{k,l,m.n=1}^d\frac{\mathcal{O}_{kl}^{(\a)}\mathcal{O}_{mn}^{(\b)}}{d^2}
+\sum_{\a=1}^L\sum_{k,l=1}^d(\mathcal{O}_{kl}^{(\a)})^{2}\\
=&L^2-L+Ld
\end{align*}
\begin{align}
\tr W_L=&d(d-1),\nonumber\\
\tr W_L^{\dagger} W_L=&(d-1)(d+L-1),
\end{align}
according to (\ref{se}),
\begin{align}
D_{sep}(\rho)\ge& -\frac{1}{\sqrt{L(d-1)}}\tr W_0\r.
\end{align}

In \cite{chruscinski2018entanglement}, the authors showed that a $3\otimes 3$ PPT state 
\begin{equation}
\r=\frac{1}{15}\begin{pmatrix}
1& 0& 0& 0& 1& 0& 0& 0& 1\\ 
0& 2& 0& 0& 0& -1& -1& 0& 0\\
 0& 0& 2& -1& 0& 0& 0& -1& 0\\ 
 0& 0& -1& 2& 0& 0& 0& -1& 0\\ 
 1& 0& 0& 0& 1& 0& 0& 0& 1\\
 0& -1& 0& 0& 0& 2& -1& 0& 0\\
 0& -1& 0& 0& 0& -1& 2& 0& 0\\
  0& 0& -1& -1& 0& 0& 0& 2& 0\\ 
  1&0& 0& 0& 1& 0& 0& 0& 1
\end{pmatrix},
\end{equation}
can be detected by the following witness $W$ which is generated by 4 MUBs,
\begin{equation}
W=\frac{1}{3}\begin{pmatrix}
4& 0& 0& 0& -1& 0& 0& 0& -1\\ 0& 1& 0& 0& 0& 2& 2& 0& 0\\0& 0& 1& 2& 0& 0& 0& 2& 0\\0& 0& 2& 1& 0& 0& 0& 2& 0\\-1& 0& 0& 0& 4& 0& 0& 0& -1\\ 0& 2& 0& 0& 0& 1& 2& 0& 0\\0& 2& 0& 0& 0& 2& 1& 0& 0\\0& 0& 2& 2& 0& 0& 0& 1& 0\\-1& 0& 0& 0& -1& 0& 0& 0& 4
\end{pmatrix},
\end{equation}
there they showed that $\tr W\rho=-\frac{2}{15}$. Then we have 
\begin{align*}
D_{sep}(\r)\ge \frac{\sqrt{2}}{30}.
\end{align*}

Next we present a second bound derived from an entanglement criterion shown in \cite{vitagliano2011spin}. Let $\{g_k|k=1,2,\cdots,d^2-1\}$ be the set of traceless $SU(d)$ generators with 
$\tr(g_kg_l)=2\delta_{kl},$
 and $N$-qudit collective $G_k$ as $G_k=\sum_{n}g_k^{(n)}$, here $$g_k^{(n)}=\underbrace{I\otimes  \cdots\otimes I}_{n-1}\otimes g_k\otimes \underbrace{I\otimes \cdots\otimes I}_{N-n}$$ is the operator acting on the $n$-th subsystem $g_k$. \par 
\indent In \cite{vitagliano2011spin}, the authors proposed the following spin-squeezing inequality,
\begin{align}
\sum_{k=1}^{d^2-1}(\triangle G_k)^2\ge 2N(d-1),\label{g1}
\end{align}
here $$\Delta G_k=\langle G_k^2\rangle-\langle G_k\rangle^2.$$

 In this manuscript, we consider the scenario when $\r_{AB}$ is a bipartite mixed state, $N=2$. Hence the inequality (\ref{g1}) can be written as
\begin{align}
&\sum_{k=1}^{d^2-1}[\langle G_k^2\rangle-\langle G_k\rangle^2 ]\nonumber\\
=&[\langle\sum_{k=1}^{d^2-1}(G_k-\langle G_k\rangle I)^2-4(d-1)\rangle]\ge 0.\label{d1}
\end{align}
Let $W_2=\sum_{k=1}^{d^2-1}(G_k-\langle G_k\rangle I)^2-4d+4,$ 
\begin{align}
&\tr(W_2^{\dagger}W_2)-(\frac{\tr W_2}{d})^2\nonumber\\
\le&144d^2-224d+112.\label{d2}
\end{align}
The inequalities $(\ref{d2})$ are due to the Lemma \ref{l1} in Sec. \ref{app}. Hence, according to (\ref{se}),
\begin{align}
D_{sep}(\rho)\ge& -\frac{\tr W_2\r}{\sqrt{144d^2-224d+112}}.
\end{align}

\subsection{The distance between a bipartite pure state and the set of separable states}\label{su1}
\begin{Theorem}\label{t1}
	Assume $\ket{\psi}$ is a bipartite pure state $\ket{\psi}=\sum_{i=0}^{d-1}\sqrt{\lambda_i}\ket{ii},$ then
	\begin{align}
	D_{sep}(\ket{\psi})=\sqrt{1-\sum_i\lambda_i^2}.
	\end{align}
\end{Theorem}

The proof of Theorem \ref{t1} is placed in Sec. \ref{app}

Based on Theorem \ref{t1}, we can obtain the following bounds of entanglement measures based on the distance $D_{sep}(\cdot).$
\begin{Theorem}\label{t2}
	Assume $\rho_{AB}$ is a bipartite mixed state,  then 
	\begin{align*}
	C(\r_{AB}) \ge& \sqrt{2}D_{sep}(\r_{AB}),\\
	E(\r_{AB})\ge &-\log(1-D_{sep}^2(\r_{AB})),\\
	E_g(\r_{AB})\ge& D_{sep}^2(\r_{AB}).
	\end{align*}
\end{Theorem}

Here we place the proof of Theorem \ref{t2} in Sec. \ref{app}.
\section{Conclusion}\label{se4}
In this paper, we have presented the lower bounds of entanglement measures for bipartite mixed states by using classes of entanglement witnesses. First, we have obtained the lower bounds of $D_{sep}(\rho)$ according to two classes of entanglement criteria. Next we have shown the analytical representations of $D_{sep}(\ket{\psi}_{AB})$ for a pure state. Based on  $D_{sep}(\ket{\psi}_{AB}),$ we have presented the lower bounds of the entanglement measures, concurrence, entanglement of formation and geometrical entanglement measures. Comparing with the method by performing the tomography of a bipartite state $\r$ to know its whole density matrix to detect and quantify the entanglement of $\r_{AB},$ the method here is much less costly. At last, the method here could be helpful to the study of multipartite entanglement theory, and we hope our work could shed some light on related studies. 
\section{Acknowledgments}
This work was supported by the National Natural Science Foundation of China (Grant No.12301580), the Fundamental Research Funds for the Central Universities (Grant No.ZY2306), and Funds of College of Information Science and Technology, Beijing University of Chemical Technology (Grant No.0104/11170044115).
 \bibliographystyle{IEEEtran}
\bibliography{ref}
\section{Appendix}\label{app}

{Theorem 1:} \emph{	Assume $\ket{\psi}$ is a bipartite pure state $\ket{\psi}=\sum_{i=0}^{d-1}\sqrt{\lambda_i}\ket{ii},$ then
	\begin{align}
	D_{sep}(\ket{\psi})=\sqrt{1-\sum_i\lambda_i^2}.
	\end{align}}

\begin{proof}
	Assume $\ket{\psi}$ is a pure state, $\ket{\psi}=\sum_i \sqrt{\lambda_i}\ket{ii}$,  then we have $(U\otimes \overline{U})\ket{\psi}\bra{\psi}(U\otimes \overline{U})^{\dagger}=\ket{\psi}\bra{\psi}$, here $U$ is a diagonal unitary matrix. Next let $\r$ be the optimal separable state, $\r=\sum_i \omega_i^A\otimes\theta_i^B$, here $\omega_i^A\in \mathcal{D}(\mathcal{H}_A),\theta_i^B\in\mathcal{D}(\mathcal{H}_B),$ then 
	\begin{align}
	D_{sep}(\rho)=&\norm{\ket{\psi}\bra{\psi}-\r}_F\nonumber\\
	= &\int_{U\in \mathbb{D}(U)}\norm{(U\otimes \overline{U})(\ket{\psi}\bra{\psi}-\r)(U\otimes\overline{U})^{\dagger}}_FdU\nonumber\\
	\ge&\norm{\int_{U\in \mathbb{D}(U)}(U\otimes \overline{U})(\ket{\psi}\bra{\psi}-\r)(U\otimes\overline{U})^{\dagger}dU}_F\nonumber\\
	=&\norm{\ket{\psi}\bra{\psi}-\int_{U\in \mathbb{D}(U)}(U\otimes\overline{U})\r(U\otimes U)^{\dagger}dU}_F,\label{s1}
	\end{align}here $\mathbb{D}(U)$ is the set of diagonal unitary operators.

	As $\r$ is the optimal, and $\int_{U\in \mathbb{D}(U)}(U\otimes\overline{U})\sigma^{'}(U\otimes\overline{U})^{\dagger}dU$ is a separable state, then 
	\begin{align}
	&D_{sep}^{'}(\ket{\psi}_{AB})\nonumber\\
	=&\norm{\ket{\psi}\bra{\psi}-\int_{U\in \mathbb{D}(U)}(U\otimes\overline{U})\r(U\otimes \overline{U})^{\dagger}dU}_F,\label{s2}
	\end{align}

	Let 
	\begin{align*}
	\vartheta=&\int_{U\in \mathbb{D}(U)} (U\otimes \overline{U})\varrho(U\otimes \overline{U})^{\dagger} dU\nonumber\\
	(\vartheta)_{ijkl}=&\int\int \int \int z_i\overline{z_j}z_l\overline{z_k}\varrho_{ijkl} dz_i dz_jdz_kdz_l\nonumber\\
	=&\left\{\begin{aligned}
	\varrho_{ijkl},\hspace{3mm} \textit{if $(i,l)=(j,k)$,}\nonumber\\
	\varrho_{ijkl},\hspace{3mm} \textit{if $(i,l)=(k,j)$,}\nonumber\\
	0,\hspace{11mm}  otherwise.
	\end{aligned}\right. 
	\end{align*}
	here $\mathbb{D}(U)$ is the set consisting of all diagnoal unitary matrices, $\varrho_{ijkl}$ denotes the coefficient of the state $\varrho$ in the basis $\ket{ij}\bra{kl}$. Hence we only need to compute the minimum of the following equality, here the minimum takes over all the positive operators $\varrho$ that can be written as $\varrho=\sum_i \omega_i^A\otimes\theta_i^B,$ $\omega_i^A$ and $\theta_i^B$ are positive operators on the systems $A$ and $B$, respectively.
	
	\begin{align}
	&\norm{\ket{\psi}\bra{\psi}-\sum_{i\ne j}\varrho_{iijj}\ket{ii}\bra{jj}-\sum_{i,j}\varrho_{ijij}\ket{ij}\bra{ij}}_F\nonumber\\
	=&\norm{\sum_{i,j}\sqrt{\lambda_i\lambda_j}\ket{ii}\bra{jj}-\sum_{i\ne j}\varrho_{iijj}\ket{ii}\bra{jj}-\sum_{i}\varrho_{iiii}\ket{ii}\bra{ii}}_F\nonumber\\+&\norm{\sum_{i\ne j}\varrho_{ijij}\ket{ij}\bra{ij}}_F\nonumber\\
	=&\norm{\sum_{i}(\lambda_i-\varrho_{iiii})\ket{ii}\bra{ii}+\sum_{i\ne j}[(\sqrt{\lambda_i\lambda_j}-\varrho_{iijj})\ket{ii}\bra{jj}]}_F\nonumber\\
	+&\norm{\sum_{i\ne j}\varrho_{ijij}\ket{ij}\bra{ij}}_F\nonumber\\
	=&\norm{\sum_{i}(\lambda_i-\varrho_{iiii})\ket{ii}\bra{ii}+\sum_{i\ne j}[(\sqrt{\lambda_i\lambda_j}-\varrho_{iijj})\ket{ii}\bra{jj}]}_F\nonumber\\+&\sqrt{\sum_{i\ne j}|\varrho_{ijij}|^2}\nonumber\\
	\ge &\norm{\sum_{i}(\lambda_i-\varrho_{iiii})\ket{ii}\bra{ii}+\sum_{i\ne j}[(\sqrt{\lambda_i\lambda_j}-\varrho_{iijj})\ket{ii}\bra{jj}]}_F\nonumber\\+&\norm{\sum_{i< j}(\varrho_{iijj}\ket{ii}\bra{jj}+\overline{\varrho_{iijj}}\ket{jj}\bra{ii})}_F\nonumber\\
	\ge&\norm{\sum_{i}(\lambda_i-\varrho_{iiii})\ket{ii}\bra{ii}+\sum_{i\ne j}\sqrt{\lambda_i\lambda_j}\ket{ii}\bra{jj}}_F\nonumber\\
	=&\norm{\ket{\psi}\bra{\psi}-\sum_i\varrho_{iiii}\ket{ii}\bra{ii}}_F\nonumber\\
	=&\sqrt{\norm{\sum_i(\lambda_i-\varrho_{iiii})\ket{ii}\bra{ii}}^2_F+\norm{\sum_{i\ne j}\sqrt{\lambda_i\lambda_j}\ket{ii}\bra{jj}}^2_F}\nonumber\\
	\ge&\norm{\sum_{i\ne j}\sqrt{\lambda_i\lambda_j}\ket{ii}\bra{jj}}_F\nonumber\\
	=&\sqrt{1-\sum_i\lambda_i^2}.
	\label{s3}
	\end{align}
	As $\varrho$ is separable, $\varrho$ satisfies the PPT condition, then $|\varrho_{ijij}|^2+|\varrho_{jiji}|^2\ge2|\varrho_{ijij}\varrho_{jiji}|\ge|\varrho_{iijj}|^2+|\overline{\varrho_{iijj}}|^2,$ hence the first inequality is valiid. The last inequality is due to the triangle inequality of the trace norm. The last equality is due to that when $\tr(A^{\dagger}B)=0,$ $\norm{A+B}_F=\sqrt{\norm{A}_F^2+\norm{B}_F^2}.$
\end{proof}

\begin{Lemma}\cite{vitagliano2011spin}\label{l1}
	Let $\{g_k|k=1,2,\cdots,d^2-1\}$ be the set of traceless $SU(d)$ generators with 
	$$\tr(g_kg_l)=2\delta_{kl},$$ 
	then
	\begin{align}
	\tr(g_kg_l)=&2\delta_{kl},\label{p1}\\
	\sum_{k=1}^{d^2-1}(g_k)^2=&2\frac{d^2-1}{d}I,\label{p2}\\
	\sum_{k=1}^{d^2-1}\langle g_k\rangle^2=&2(\tr\rho^2-\frac{1}{d}),\label{p3}\\
	\sum_{k=1}^{d^2-1}g_k\otimes g_k=&2(F-\frac{1}{d}I).\label{p4}
	\end{align}
\end{Lemma}

Theorem 2: \emph{	Assume $\rho_{AB}$ is a bipartite mixed state,  then }
	\begin{align*}
	C(\r_{AB}) \ge& \sqrt{2}D_{sep}(\r_{AB}),\\
	E(\r_{AB})\ge &-\log(1-D_{sep}^2(\r_{AB})),\\
	E_g(\r_{AB})\ge& D_{sep}^2(\r_{AB}).
	\end{align*}

\begin{proof}
	Assume $\rho_{AB}$ is a bipartite mixed state, $\{p_i,\ket{\psi_i}\}$ is the optimal decomposition of $\r_{AB}$ in terms of $C_{AB}$,
	\begin{align}
	C(\rho_{AB})=&\sum_i p_i C(\ket{\psi_i})\nonumber\\
	=&\sum_i p_i \sqrt{2(1-\tr(\rho_i^A)^2}\nonumber\\
	=&\sum_i p_i\sqrt{2(1-\sum_k{\lambda^{(i)}_k}^2)}\nonumber\\
	=&\sum_i p_i\sqrt{2}D_{sep}(\ket{\psi_i})\nonumber\\
	\ge&\sqrt{2}D_{sep}(\rho_{AB}), \label{ce}
	\end{align}
	
	In the first equality $\{p_i,\ket{\psi_i}\}$ is the optimal decomposition of $\r_{AB}$ in terms of $C$, $\{\lambda^{(i)}_k|k=0,1,2,\cdots, d-1\}$ is the set consisting of the Schmidt vectors of $\ket{\psi_i}$ in the second equality, the third equality is due to Theorem \ref{t1}.
	
	Next assume $\rho_{AB}$ is a bipartite mixed state, $\{p_i,\ket{\psi_i}\}$ is the optimal of $\r_{AB}$ in terms of $E(\r_{AB})$,
	\begin{align}
	E(\r_{AB})=&\sum_i p_iE(\ket{\psi_i})\nonumber\\
	=&\sum_i p_i S(\rho^{(i)}_A)\nonumber\\
	\ge&-\sum_i p_i\log(1-D_{sep}^2(\ket{\psi^{(i)}})\nonumber\\
	\ge&-\log(1-D^2_{sep}(\rho)).\label{ef}
	\end{align}
	In the first equality, we assume $\{p_i,\ket{\psi_i}\}$ is the optimal decomposition of $\r_{AB}$ in terms of $E$. As $S(\rho)\ge \frac{1}{1-\b}\log(\rho^{\beta})$ when $\b>1$ \cite{song2016}, the third inequality is valid. The last inequality is due to the convexity of $-\log(1-x^2).$\par 
	When $\{p_i,\ket{\psi_i}\}$ is the optimal decomposition of $\r_{AB}$ in terms of $E_g(\r(AB))$,
	\begin{align}
	E_g\r_{AB})=&\sum_i p_iE_g(\ket{\psi_i})\nonumber\\
	\ge&\sum_i p_iD_{sep}^2(\ket{\psi^{(i)}})\nonumber\\
	\ge&D^2_{sep}(\rho).
	\end{align}
	In the first equality, we assume $\{p_i,\ket{\psi_i}\}$ is the optimal decomposition of $\r_{AB}$ in terms of $E_g$. As $\norm{M^{p}}^{\frac{1}{p}}$ for positive $M$ is monotone decreasing, the third inequality is valid. The last inequality is due to the convexity of $x^2.$
\end{proof}
\end{document}